\documentstyle[aps,eqsecnum,preprint]{revtex} \baselineskip 12pt
\newcommand{\be}{\begin{equation}}
\newcommand{\ee}{\end{equation}}
\newcommand{\bea}{\begin{eqnarray}}
\newcommand{\eea}{\end{eqnarray}}
\begin{document}
\title{ A PT-Invariant Potential With Complex QES Eigenvalues} 

\vspace{.4in}

\author{Avinash Khare }

\address{
Institute of Physics, Sachivalaya Marg,\\ Bhubaneswar-751005, India,\\
Email:  khare@iopb.res.in}

\author{ Bhabani Prasad Mandal}

\address{
 S.N. Bose National Center for Basic Sciences,\\
Block-JD; Sector-III; Salt Lake ,\\
 Calcutta-700 091, India.\\
Email: bpm@boson.bose.res.in}

\vspace{.4in}

\maketitle

\vspace{.4in}

\begin{abstract}

We show that the quasi-exactly solvable eigenvalues of the 
Schr\"odinger equation for the PT-invariant 
potential $V(x) = -(\zeta \cosh 2x -iM)^2$ 
 are complex conjugate pairs in case the parameter M is an even integer
while they are real in case M is an odd integer. 
We also show that whereas the PT symmetry is spontaneously broken in
the former case, it is unbroken in the latter case.
\end{abstract}

\newpage

In non-relativistic quantum mechanics, one usually chooses a real 
(Hermitian) potential so as to ensure the real energy eigenvalues of the
corresponding Schr\"odinger equation. Few 
years ago,  
Bender and others \cite{bed,oth} have 
studied several complex potentials
which are invariant under the combined symmetry $PT$ and showed that 
even in all these cases, the energy eigenvalues of the Schr\"odinger equation
are real so long as the $PT$ symmetry is not spontaneously broken.
This seem to suggest that instead of Hermiticity, it may be  
enough to have PT-invariant Hamiltonian so as to have real energy 
eigenvalues. It must however be noted that so far, this is merely a conjecture
which is being supported by several examples. Of course, even if this conjecture
is true, there are several basic questions which will have to be  
addressed before one can take the PT-invariant potentials more seriously.

It must however be noted here that PT invariance is {\it not sufficient} to
ensure the reality of spectrum. As has been noted by Bender and Boettcher 
\cite{bed}, one has to ensure that the PT symmetry
is not spontaneously broken. As an illustration, using extensive numerical and 
asymptotic studies, they have shown that the eigenvalues of the potential
$-(ix)^{N}, (N$ real) are real if $N \ge 2$ while there are finite number
of real and infinite number of complex conjugate pairs of eigenvalues if
$1 < N < 2$. They also showed that whereas PT symmetry is unbroken if
$N \ge 2$, it is spontaneously broken if $1 < N < 2$.  

Thus it is clearly of great importance 
to understand as to when the PT symmetry is
spontaneously broken and when it remains unbroken. 
As a first step in that direction, it may be worthwhile to look for some
analytically solvable PT-invariant potentials and  
try to understand the spontaneous breaking and nonbreaking of the PT symmetry. 
The purpose of this note is to study one such example. 
In particular, we consider the system described by the
non-Hermitian but PT-invariant Hamiltonian ($\hbar =2m =1$)
\be\label{1}
H= p^2 - (\zeta\cosh 2x -iM)^2 \, ,
\ee
where the parameter $\zeta $ is real and parameter $M$ has only
integer values. 
We show that the quasi-exactly solvable (QES) 
eigen values \cite{ush} of 
this Hamiltonian are complex conjugate pairs 
in case the parameter M is an even integer and that in this case the
PT symmetry is indeed spontaneously broken. On the other hand, when
$M$ is an odd integer then the QES eigenvalues of this Hamiltonian
are real and precisely in this case the PT symmetry remains unbroken.

Let us first show that the above potential is PT-invariant. 
To that end, notice that
under T reflection one replaces $i$ by $-i$ while under P reflection one 
replaces
 $x$ by $a-x$, where $a/2$ is the origin about which one is performing the
 parity reflection. It is then easily checked that the Hamiltonian (1) 
 is PT symmetric under the parity reflection $x \rightarrow i\pi/2 -x$.

We substitute
\be\label{aa}
\psi(x) = e^{i\frac{ \zeta}{2}\cosh 2x}\phi(x) \, ,
\ee
in the Schro$\ddot{o}$dinger equation $H\psi= E\psi$ with $H$ as given by
Eq. (\ref{1}) and obtain
\begin{equation}
\phi^{\prime\prime}(x) +2i\zeta\sinh 2x\phi^\prime (x) + \left
 [(E-M^2+\zeta^2) -2i(M-1)\zeta\cosh 2x \right ]\phi(x) =0 \, .
\label{1st} 
\end{equation}
Ordinarily, the boundary conditions that give the quantized energy levels are
$\psi (x) \rightarrow 0$ as $\mid x \mid \rightarrow \infty$ on the real axis.
However, in the present 
case, we have to continue the eigenvalue problem into the 
complex-x plane \cite{bt}. 
On putting, $x = u+iv$ where $u,v$ are real, it is easy to 
see that for $u > 0$ the above boundary condition is satisfied so long as 
$ -\pi < v < -\pi/2 \ (mod\ \pi$) while for $u < 0$ it is satisfied if
$-\pi/2 < v <0 \ (mod \ \pi$). 

On further substituting \cite{km}
\begin{equation}
z=\cosh 2x -1;\ \ \ 
\phi = z^s\sum_{n=0}^\infty \frac{ R_n(E)}{n!} 
\left ( \frac{ z+2}{2} \right )^{ \frac{ n}{2}} \, ,
\label{2nd}
\end{equation}
we obtain the three-term recursion relation $(n\ge 0)$
\begin{eqnarray} 
R_{n+2}(E)&-&\left [ n^2 +4(s-i\zeta)n + 4s^2 + \left .
E-M^2+\zeta^2+2i(M-1)\zeta \right .  \right ]R_n(E) \nonumber \\  
+&&4i\zeta \left [ M+1-2s-n \right ]n(n-1) R_{n-2}(E)=0 \, ,
\label{r1}
\end{eqnarray}
provided $2s^2 = s$ i.e. either $s=0$ or
$s= \frac{ 1}{2}$. Thus we have two sets of independent solutions; one for
$s=0$ and other for $s= \frac{ 1}{2}$. Note that the parameter $s$ is not
contained in the potential and this is perhaps related to the fact that 
for any (integer) M $(>1)$, the QES solutions corresponding to both even and odd
number of nodes are obtained.  

From Eq. (\ref{r1}) we observe that the even and odd
polynomials $R_n(E)$ do not mix with each other and hence 
we have two separate three-term recursion relations
depending on whether $n$ is odd or even. In particular, it is easily 
shown that the three-term recursion relations corresponding to the even and odd 
$n$ cases respectively are given by $(n\ge 1)$
\begin{eqnarray}
P_n(E)&-& \left [ 4n^2 +8n(s-i\zeta-1)+4s^2-8s +4+6i\zeta +E -(M-i\zeta)^2
\right ]P_{n-1}(E) \nonumber \\   
&+& 8i\zeta(n-1)(2n-3)\left ( M+3-2s-2n \right ) P_{n-2} = 0 \, ,
\label{r2}  \\  
Q_n(E)&-& \left [ 4n^2 +4n(2s-2i\zeta-1)+4s^2-4s +1+2i\zeta +E
-(M-i\zeta)^2
\right ]Q_{n-1}(E) \nonumber \\  
&+& 8i\zeta(n-1)(2n-1)\left ( M+2-2s-2n \right ) Q_{n-2} = 0  \, , 
\label{rr1}
\end{eqnarray}
with $P_0(E)=1, Q_0(E)=1. $

We have studied the properties of the weakly orthogonal polynomials
$P_n (E)$ and $Q_n (E)$ and we find that many of these are similar 
to the Bender-Dunne polynomials \cite{bd}. Further,
we find that for integer values of $M$, there are $M$ QES 
eigen values and they are complex conjugate pairs 
if $M$ is an even integer
(M=2,4,...) while they are real if $M$ is an odd integer 
(provided $|\zeta|\le \zeta_{c}$,
where $\zeta_{c}$ is a function of $M$).
In particular, if $M = 2k+2 (2k+1)$
($k=0,1,...$), then $k+1$ levels  
are obtained from the zeros of the
orthogonal polynomial $P_{k+1} (E)$ while $k+1 (k$) levels are 
obtained from the zeros of $Q_{k+1} (Q_k)$ which we term
as the critical polynomials. 
Further, all higher $P_n $ and $ Q_n$ polynomials exhibit factorization 
property.

Let us first consider few even integer values of $M$ and show explicitly
that the QES eigenvalues are complex conjugate pairs and further the PT
symmetry is spontaneously broken. We start from M=2 for which
case the critical polynomials $P_1 (E)$ (with $s=1/2$) and $Q_1 (E)$ 
(with $s=0$) are given by 
\be\label{2}
P_1 (E) =  E -3 + \zeta^2 + 2i \zeta \, , \ \ Q_1 (E) = E -3 +\zeta^2 -2i \zeta \, .
\ee
hence the two QES complex conjugate pair of energy eigenvalues are 
\be\label{3}
E_{\pm} = 3 - \zeta^2 \pm 2i \zeta \, .
\ee
The corresponding eigenfunctions are as given by eq. (\ref{aa}) with
\be\label{3a}
\phi_{+} = \cosh x \, , \ \phi_{-} = \sinh x \, .
\ee
Now under the PT transformation as defined above, $\cosh x \rightarrow 
-i\sinh x, \sinh x \rightarrow i\cosh x$ while $i\cosh 2x$ remains
invariant so that the two wave functions are not invariant under PT, i.e.
in this case (where the eigenvalues are complex conjugate pair), the PT
symmetry is spontaneously broken.

For $M =4$, the critical polynomials are
\be\label{4}
P_2 (E) = \epsilon^2 +(10+ 4i\zeta)\epsilon +3(4\zeta^2 +12i\zeta +3) \, ,
\ee
\be\label{5}
Q_2 (E) = \epsilon^2 +(10 -
4i\zeta)\epsilon +3(4\zeta^2 -12i\zeta +3) \, ,
\ee  
where $\epsilon = E - 16 +\zeta^2$. Hence the four (two complex conjugate
pair of) 
eigenvalues are
\be\label{6}
E^{1}_{\pm} = 11 -\zeta^2 -2i \zeta \pm 4\sqrt{1-i\zeta -\zeta^2} \, ,
\ee
\be\label{7}
E^{2}_{\pm} = 11 -\zeta^2 +2i \zeta \pm 4\sqrt{1+i\zeta -\zeta^2} \, .
\ee
It is easily checked that the corresponding eigenfunctions are
\be
\phi^{1}_{\pm} = A \sinh 3x + B \sinh x \, , \ 
\frac{B}{A} = \frac{E-7+\zeta^2}{2i\zeta} \, ,
\ee
\be
\phi^{1}_{\pm} = A \cosh 3x + B \cosh x \, , \ 
\frac{B}{A} = \frac{E-7+\zeta^2}{2i\zeta} \, ,
\ee
so that in this case also the PT symmetry is spontaneously broken.
It is easily checked that this is also true for all higher even 
integer values of M.

On the other hand, 
if $M$ is an odd integer, then the eigenvalues are real provided 
$\zeta^2 \le \zeta_c^2$. For example, for M=1, the critical polynomial is 
$P_1 (E) = E -1 +\zeta^2$ so that the QES level is at $E = 1 - \zeta^2$. 
In this case the eigenfunction is
\be
\phi = constant \, ,
\ee
so that the corresponding $\psi$ is indeed an eigenfunction of PT with 
eigenvalue 1 and hence PT symmetry remains unbroken.

For $M=3$, the critical polynomials are
\be 
P_2 (E) = \epsilon^2 +4\epsilon +16\zeta^2 \, , \ \ Q_1 (E) = \epsilon +4 \, ,
\ee
where $\epsilon = E -9 +\zeta^2$. Thus the three QES eigenvalues are
\be
E = 5 - \zeta^2 \, , \ \ E_{\pm} = 7 -\zeta^2 \pm 2\sqrt{1-4\zeta^2} \, .
\ee
These energy levels are real if 
$\zeta^2 \le \zeta_c^2 = 1/4$ and at $\zeta = \zeta_c$, the
two highest QES levels are degenerate.
The corresponding eigenfunctions are
\be
\phi = \sinh 2x \, , \ \phi_{\pm} = A \cosh 2x + iB
\ee
where
\be
\frac{B}{A} = \frac{4\zeta}{E-9+\zeta^2} \, .
\ee
It is easily checked that the corresponding wavefunction $\psi$ 
is indeed an eigenfunctions of PT with eigenvalue 1 while the other two 
wavefunctions ($\psi_{\pm}$) are eigenfunctions with eigenvalue -1. 
Thus the PT symmetry remains unbroken. In fact, it is easily checked that
this is indeed so for all higher odd integer values of M \cite{4019}.

It is worth pointing out that if following Finkel et al. \cite{fin}, we
change the variable, $z = e^{2x}$ and make the gauge transformation
\be
\mu (z) = z^{\frac{1-M}{2}} e^{\frac{i\zeta}{4} (z+\frac{1}{z})} \, ,
\ee
in the Hamiltonian (1) to map it to a differential operator $H_g$ (called
as gauged Hamiltonian), then this gauged Hamiltonian can be expressed in terms
of the generators of the $Sl(2,R)$ by
\be
H_g (z) = -4J_0^2 +2i\zeta (J_- - J_{+}) -c^{*} \, , 
\ee
where $c^{*} = \zeta^2 -M^2$ while the generators $J_{+,-,0}$ are given by
\be
J_{+} = \frac{\partial}{\partial z} \, , \ 
J_{0} = z\frac{\partial}{\partial z} -\frac{n}{2} \, ,    
\ J_{-} = z^2 \frac{\partial}{\partial z} -nz \, .
\ee
Further, in this case the norms and the weight functions of the corresponding
orthogonal polynomials are real. 
Of course, as expected,  
the QES eigenvalues remain unchanged. 

Finally, using the anti-isospectral
transformation of Krajewska et al. \cite{kra}, it follows that the
QES eigenvalues and eigenfunctions 
of the PT-invariant potential (with parity reflection
$\theta \rightarrow \pi/2 -\theta$) 
\be
V(\theta) = (\zeta \cos 2\theta -iM)^2 \, ,
\ee
are related to those of  the potential in Eq. (1) by
\be
\bar{E}_k = -E_{M-1-k} \, , \bar{\psi}_k (\theta) = \psi_{M-1-k} (ix) \, .
\ee 
In particular, the QES eigenvalues of this periodic potential  
are real if M is an odd integer and further it is easily shown that the
PT symmetry remains unbroken in this case. On the other hand, when M
is an even integer, then the eigenvalues are complex conjugate pairs 
and PT symmetry is spontaneously broken.
But in any case, the eigenstates are unacceptable in case M is an even 
integer \cite{km}
since the corresponding eigenfunctions do not satisfy the periodicity
property of the above potential.

It would be worthwhile to explore as to when is PT symmetry likely to
remain unbroken and when is it spontaneously broken.

{\bf Acknowledgement:} We are grateful to the referee Prof. Carl M. Bender
for pointing out the
PT-invariance of the potentials discussed in this paper and reminding about
the connection between the reality of the spectrum and the PT symmetry 
being unbroken.

\end{document}